\documentclass[a4paper,fleqn,usenatbib,useAMS]{mnras}

\usepackage{graphicx}
\usepackage{natbib}
\usepackage{color}
\usepackage{subfigure}
\usepackage{amsmath}

\def\hi{\relax \ifmmode {\mbox H\,{\scshape i}}\else H\,{\scshape i}\fi}

\def\hii{\relax \ifmmode {\mbox H\,{\scshape ii}}\else H\,{\scshape ii}\fi}

\def\nii{\relax \ifmmode {\mbox N\,{\scshape ii}}\else N\,{\scshape ii}\fi}

\def\oi{\relax \ifmmode {\mbox O\,{\scshape i}}\else O\,{\scshape i}\fi}

\def\oii{\relax \ifmmode {\mbox O\,{\scshape ii}}\else O\,{\scshape ii}\fi}

\def\oiii{\relax \ifmmode {\mbox O\,{\scshape iii}}\else O\,{\scshape iii}\fi}

\def\cii{\relax \ifmmode {\mbox C\,{\scshape ii}}\else C\,{\scshape ii}\fi}

\def\ciii{\relax \ifmmode {\mbox C\,{\scshape iii}}\else C\,{\scshape iii}\fi}

\def\civ{\relax \ifmmode {\mbox C\,{\scshape iv}}\else C\,{\scshape iv}\fi}

\def\hei{\relax \ifmmode {\mbox He\,{\scshape i}}\else He\,{\scshape i}\fi}

\def\heii{\relax \ifmmode {\mbox He\,{\scshape ii}}\else He\,{\scshape ii}\fi}

\def\mgii{\relax \ifmmode {\mbox Mg\,{\scshape ii}}\else Mg\,{\scshape ii}\fi}

\def\sii{\relax \ifmmode {\mbox S\,{\scshape ii}}\else S\,{\scshape ii}\fi}

\def\neiii{\relax \ifmmode {\mbox Ne\,{\scshape iii}}\else Ne\,{\scshape iii}\fi}

\def\ariv{\relax \ifmmode {\mbox Ar\,{\scshape iv}}\else Ar\,{\scshape iv}\fi}

\def\ni{\relax \ifmmode {\mbox N\,{\scshape i}}\else N\,{\scshape i}\fi}

\def\ariii{\relax \ifmmode {\mbox Ar\,{\scshape iii}}\else Ar\,{\scshape iii}\fi}

\def\caii{\relax \ifmmode {\mbox Ca\,{\scshape ii}}\else Ca\,{\scshape ii}\fi}

\title[]{Spectroscopic characterisation of the stellar content of ultra diffuse galaxies} 

\author[T. Ruiz-Lara et al.]{T. Ruiz-Lara,$^{1, 2}$\thanks{E-mail: tomasruizlara@gmail.com (TRL)}  M.A. Beasley,$^{1, 2}$  J. Falc\'on-Barroso,$^{1,2}$ J. Rom\'an,$^{1, 2}$ F. Pinna,$^{1,2}$  \newauthor C. Brook,$^{1, 2}$ A. Di Cintio,$^{1, 2}$ I. Mart\'in-Navarro$^{3, 4}$ I. Trujillo,$^{1, 2}$ and A. Vazdekis$^{1, 2}$ \\
$^{1}$ Instituto de Astrof\'isica de Canarias, Calle V\'ia L\'actea s/n, E-38205 La Laguna, Tenerife, Spain \\
$^{2}$ Departamento de Astrof\'isica, Universidad de La Laguna, E-38200 La Laguna, Tenerife, Spain \\
$^{3}$ University of California Observatories, 1156 High Street, Santa Cruz, CA 95064, USA \\
$^{4}$ Max-Planck Institut f\"ur Astronomie, Konigstuhl 17, D-69117 Heidelberg, Germany 
}

\begin{document}

\date{Accepted 2018 April 26. Received 2018 April 16; in original form 2018 March 16}

\pagerange{\pageref{firstpage}--\pageref{lastpage}} \pubyear{2018}

\maketitle

\label{firstpage}

\begin{abstract}

  Understanding the peculiar properties of Ultra Diffuse Galaxies (UDGs) via spectroscopic analysis is a challenging task requiring very deep observations and exquisite data reduction. In this work we perform one of the most complete characterisations of the stellar component of UDGs to date using deep optical spectroscopic data from OSIRIS at GTC.
  We measure radial and rotation velocities, star formation histories (SFH) and mean population parameters, such as ages and metallicities, for a sample of five UDG candidates in the Coma cluster.   From the radial velocities, we confirm the Coma membership of these galaxies.
  We find that their rotation properties, if detected at all, are compatible with dwarf-like galaxies. The SFHs of the UDG are dominated by old ($\sim$ 7 Gyr), metal-poor ([M/H] $\sim$ --1.1) and $\alpha$-enhanced ([Mg/Fe] $\sim$ 0.4) populations followed by a smooth or episodic decline which halted $\sim$ 2 Gyr ago, possibly a sign of cluster-induced quenching. We find no obvious correlation between individual SFH shapes and any UDG morphological properties. The recovered stellar properties for UDGs are similar to those found for DDO~44, a local UDG analogue resolved into stars. We conclude that the UDGs in our sample are extended dwarfs whose properties are likely the outcome of both internal processes, such as bursty SFHs and/or high-spin haloes, as well as environmental effects within the Coma cluster. 

\end{abstract}


\section{Introduction}

The study of the lowest surface brightness tail of the observed galactic population is essential to acquire a complete picture of galaxy formation and evolution. Particularly puzzling is the observation of extremely faint and extended systems, known since the 1980's \citep[][]{1984AJ.....89..919S, 1988ApJ...330..634I, 1988AJ.....96.1520F, 1991ApJ...376..404B, 1997AJ....114..635D}. Dynamical arguments, such as resistance against cluster-induced tidal effects, place them as one of the objects with the highest mass-to-light ratios in the Universe \citep[][]{1991ApJ...376..404B, 2015ApJ...798L..45V}, suggesting they should be dark matter dominated \citep[][]{2009MNRAS.393.1054P}.

Currently renamed as ``Ultra-Diffuse Galaxies'' \citep[UDGs, $\mu_g(0)$~$\ge$~24~mag/arcsec$^2$ and R$_{\rm eff}$~$\ge$~1.5 kpc;][]{2015ApJ...798L..45V}, interest in these systems has been recently revived. Deep photometric observations reveal large numbers of UDGs in groups and clusters of galaxies \citep[e.g.][]{2015ApJ...807L...2K, 2015ApJ...809L..21M, 2015ApJ...813L..15M, 2016A&A...590A..20V, 2016AJ....151...96M, 2017ApJ...836..191T, 2017MNRAS.468..703R, 2017MNRAS.468.4039R}, highlighting the ubiquity of this population. But, how could galactic systems acquire such observed characteristics?

\citet[][]{2015ApJ...798L..45V} suggested the possibility that UDGs are ``failed'' Milky Way-like galaxies whose star formation suffered an early quenching after losing their gas content via ram pressure stripping \citep[][]{2015MNRAS.452..937Y}. However, several works focused on the atomic gas (\hi) properties of blue UDGs \citep[][]{2017ApJ...836..191T, 2017arXiv171006557S} or the dynamics \citep[][]{2016ApJ...819L..20B} and the size  \citep[][]{2016ApJ...830...23B, 2016ApJ...822L..31P} of the globular cluster system of UDGs have found masses more compatible with those of dwarfs, with the possible exceptions of DF44 and DFX1 \citep[][]{2016ApJ...828L...6V, 2017ApJ...844L..11V, 2017MNRAS.464L.110Z}. \citet[]{2016MNRAS.459L..51A} found that dwarf galaxies embedded in high angular momentum haloes can produce the UDG properties \citep[see also][]{2017MNRAS.470.4231R}. \citet[][]{2017MNRAS.466L...1D} proposed that episodic, bursty star formation in dwarf galaxies could also lead to the large sizes of UDGs as feedback-driven outflows could cause the expansion of the stellar and dark matter components in such systems \citep[see also][]{2017arXiv171104788C}. Both models predict that UDGs should be found both in the field as well as in clusters.

A key to understanding the real nature of UDGs might be imprinted in their stellar populations, although such analysis is hindered by their low surface brightness. To date, most of the works have relied on photometric information to describe the stars populating UDGs \citep[][]{2015ApJ...798L..45V, 2017MNRAS.468.4039R, 2017MNRAS.468..703R, 2017ApJ...836..191T, 2017arXiv171105272P}, although lately more effort has been put towards a reliable spectroscopic characterisation of UDGs \citep[][]{2017ApJ...838L..21K, 2017arXiv170907003G}. Currently, the growing view is that UDGs are mainly red, metal-poor, old systems with the existence of a secondary population of blue, low surface brightness systems located in low-density environments, hinting a possible evolution from blue-UDGs to red-UDGs \citep[][]{2017MNRAS.468.4039R, 2017arXiv170904474G}.

In order to gain insight into the build-up and subsequent evolution of UDGs, we need detailed information of their star formation histories (SFHs). \citet[][]{2018arXiv180109695F} have recently presented such an analysis for 7 UDGs. These authors found intermediate luminosity-weighted ages ($\sim$~7~Gyr), low metallicities for their stellar masses ([Z/H]~$\sim$~-0.7) and enhanced [Mg/Fe] values consistent with the properties of the general dwarf galaxy population. Based on this information and the UDG kinematics in the Coma cluster \citep[from][]{2018arXiv180109686A}, they conclude that multiple formation pathways are compatible to explain UDG properties.

In this contribution, we present a thorough characterisation of the stellar kinematics as well as SFHs from the analysis of deep spectroscopy of 5 UDG candidates in the Coma cluster. We use the inversion code {\tt STECKMAP} \citep[STEllar Content and Kinematics via Maximum A Posteriori likelihood;][]{2006MNRAS.365...74O, 2006MNRAS.365...46O}. The combination of spectroscopic data of the highest quality and state-of-the-art full spectral fitting techniques will provide unprecedented clues on how UDGs come to be.

The paper is structured as follows. We describe the data, sample, and methodology in Sects.~\ref{data} and~\ref{analysis}. The main results concerning the stellar content of UDGs as well as their possible rotation are given in Sects.~\ref{stars} and~\ref{kin}. In Sect.~\ref{discussion}, we discuss the main implications of these results on the nature of UDGs as well as their formation and evolution. A summary and the main conclusions can be found in Sect.~\ref{conclusions}. We adopt a cosmology with H$_0$ = 69.7 km s$^{-1}$ Mpc$^{-1}$, $\Omega_m$ = 0.281 and $\Omega_\Lambda$ = 0.719 \citep[][]{2013ApJS..208...19H}, resulting in a cosmological scale of 0.473 kpc/arcsec \citep[][]{2016ApJS..225...11Y} for Coma \citep[cz = 7100 kms$^{-1}$,][]{2000ApJ...533..125K}.

\section{Data and sample characterization}
\label{data}

\begin{figure*}
\centering 
\includegraphics[width = 0.95\textwidth]{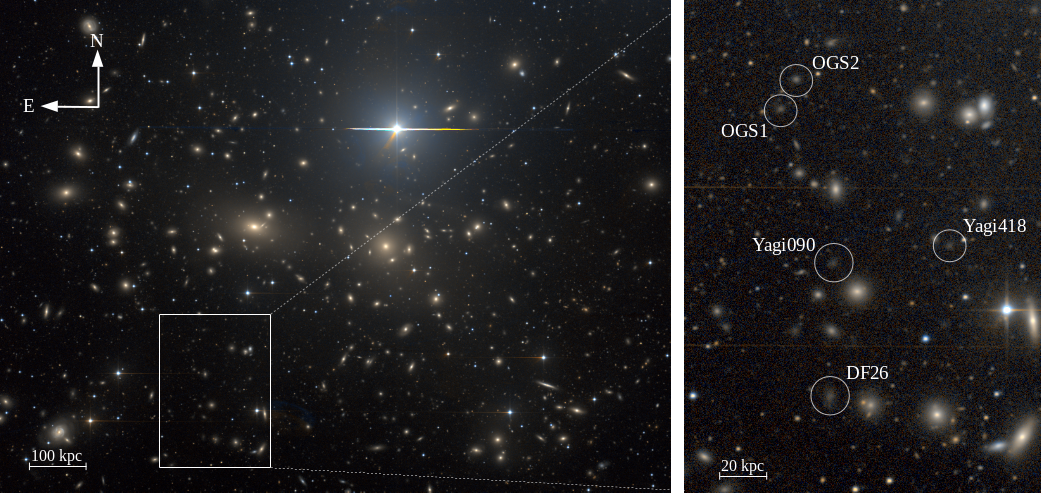} \\   
\caption{Composite image of the Coma cluster using B- and V-band observations from the OGS telescope (see text for details). The zoom-in rectangle shows the targets selected for the deep GTC spectroscopic observations.} 
\label{coma_sketch} 
\end{figure*}

We analyse deep spectroscopic data of a region within the Coma cluster using the OSIRIS imager and spectrograph\footnote{Complete information regarding the OSIRIS instrument can be found at \url{http://www.gtc.iac.es/instruments/osiris/}} mounted at the Gran Telescopio Canarias, in the Observatorio del Roque de los Muchachos. Such region was chosen in order to maximise the number of UDG candidates within the OSIRIS field of view (see Fig.~\ref{coma_sketch}). These candidates were selected from the catalogues of \citet[][]{2015ApJ...798L..45V} and \citet[][]{2015ApJ...807L...2K}, and later characterised in \citet[][]{2016ApJS..225...11Y}. The observations were part of the \verb|GTC63-16A| program and were carried out in dark nights from March-2016 to May-2017. In total, 20 hours divided into 40 exposures of 1800s were taken using the OSIRIS MOS configuration to obtain spectroscopic data for 9 UDG candidates (position angle of $\sim$ 270$^{\rm o}$). We combine the R2000B grism with a slit width of 1.2'' to have a spectral resolution (FWHM) of $\sim$ 4.5 \AA~over the wavelength range from 3950~\AA~to 5700~\AA. Given the expected redshift of the analysed objects (Coma), this wavelength range would cover the more relevant spectroscopic absorption features for a proper characterisation of the stellar content of the targets under analysis.

Standard MOS-data reduction steps such as bias subtraction, flat fielding, slits extraction, C-distortion correction, wavelength calibration, and cosmic rays removal \citep[L.A. Cosmic,][]{2001PASP..113.1420V} were performed making use of a reduction pipeline specifically designed for dealing with OSIRIS-MOS data based on a set of Python-IDL routines. Special care was taken in the sky subtraction step due to the low surface brightnesses of UDGs (more than three magnitudes fainter than the brightness of the sky during a moonless night in La Palma, $\rm \mu_V\sim$ 21.9 $\rm mag/arcsec^{2}$, \citealt[][]{1998NewAR..42..503B}). For that, we make use of the sky subtraction algorithm described in \citet[][]{2003PASP..115..688K}. After characterising the CCD distortions and curvature of the spectral features (C-distortion and wavelength calibration), we obtain the characteristic sky spectrum in each individual slit from those pixels free of any light contamination.

After reducing all individual observed frames (40 per slit position), we performed two different extractions in order to obtain the final spectra to be analysed (three per candidate/slit). On the one hand we combine all the pixels around the position where the candidates were placed and brighter than the $\sim$ 60$^{\rm th}$ percentile of the slit light distribution. The lack of bright sources (foreground stars or background galaxies) allow us to claim that all the light included in the computation of these spectra is coming from the UDG candidates. This first approach allows us to obtain a single spectrum of the highest possible S/N for each galaxy in order to study their stellar properties as well as their star formation histories (S/N per $ \rm \AA$~ranging from 17 to 46, see Fig.~\ref{fits_plot}). Hereafter we will denote this first spectrum as spectrum ``W''. On the other hand, we extract two more spectra, using the same criterium but avoiding this time the central part of each galaxy (pixels brighter than the $\sim$ 95$^{\rm th}$ percentile). In this second approach, although the S/N is significantly lower as to reliably obtain information about its stellar content, we can use these two spectra to try to infer some information regarding the possible rotation curve of the objects under analysis (see Sect.~\ref{kin}). We will name these two spectra as spectrum ``L'' (pixels in the left-hand part of the slit) and spectrum ``R'' right-hand part of the slit). Given the geometry of the observations and the position angle of the individual slits, spectra ``L'' and ``R'' cover the southern and northern parts of the galaxies, respectively.

\begin{table*}
{\normalsize
\centering
\begin{tabular}{ccccc}
\hline\hline
Galaxy & RA & Dec. & $\rm \langle \mu_{B} \rangle$ & R$_{\rm eff}$ \\ 
 &  (h:m:s) & ($^{\rm o}$ $'$ $''$) &  (mag/arcsec$^2$) &  ($''$) \\ 
$(1)$ & $(2)$ & $(3)$ & $(4)$ & $(5)$ \\ \hline
 DF26*            & 13:00:20.6 & +27 47 12.3  &  25.6  &   5.1   \\
 Yagi~090*        & 13:00:20.4 & +27 49 24.0  &  25.9  &   3.4   \\
 Yagi~418*        & 13:00:11.7 & +27 49 41.0  &  25.5  &   2.8   \\
 OGS1**           & 13:00:24.3 & +27 51 55.5  &  25.2  &   3.2   \\
 OGS2**           & 13:00:23.2 & +27 52 24.6  &  24.4  &   2.7   \\ \hline
\end{tabular}
\caption{Sample of galaxies. (1) Name given in this work to each of the analysed galaxies; (2) right ascension (J2000); (3) declination (J2000); (4) Surface brightness (B-band) averaged over 1 effective radius; (5) Effective radius (arcsec). (*) Structural decomposition from \citet[][]{2016ApJS..225...11Y}; (**) structural decomposition from our OGS data (see text for details).} 
\label{galaxy_tab}
}
\end{table*}

The sky subtraction proved unsatisfactory in 4 out of the 9 cases analysed, resulting thus in a sample composed by 5 UDG candidates. The deficient sky subtraction in these four galaxies is the outcome of the combination of low surface brightness of the galaxies and, specially, their location on the CCDs of the OSIRIS detector. It is known that the efficiency of the OSIRIS detector slightly changes along its 2D extension, changes that turn out crucial when dealing with such faint objects. In this work, sky subtraction is performed in each of the individual exposures, where the collected signal is lower than the uncertainties in the sky determination and those induced by the variability in the OSIRIS detector. As a consequence, despite the amount of individual exposures at our disposal, the poor sky subtraction performed in the individual frames hampered the extraction of high-quality integrated spectra. We conclude that not only large telescopes are needed to carry out spectroscopic analysis of UDGs, but stable detectors and exquisite sky subtraction techniques. We should highlight here that the rejected galaxies do not share any particular characteristics in terms of surface brightness, effective radius, or colour, discarding any possible observational bias in the results of this paper.

Table~\ref{galaxy_tab} summarises the main characteristics of the final sample of galaxies analysed in this work. Three of these systems (DF26, Yagi~090, and Yagi~418) were previously identified as UDGs and their structural parameters determined \citep[][]{2015ApJ...798L..45V, 2016ApJS..225...11Y}. In addition, the stellar content of DF26 and Yagi~418 have also been analysed previously in \citet[][]{2018arXiv180109695F}, allowing for an interesting comparison. The other two (OGS1 and OGS2) were visual candidates within the observed OSIRIS field not previously characterised. For the latter, we determined their morphological parameters from deep photometric follow-up observations with the Optical Ground Station (OGS) telescope in the Iza\~na Observatory, Tenerife, Spain. We observed 1.5 hours in B and V bands reaching 28.4 and 28.2 $\rm mag/arcsec^{2}$ (3$\sigma$ measured in 10$\times$10 arcsec boxes), respectively. This is a similar depth to the date presented in \citet[][]{2015ApJ...798L..45V}. The final structural parameters were obtained using {\tt Imfit} \citep[][]{2015ApJ...799..226E} with a single Sersic component, in a similar way as done by \citet[][]{2016ApJS..225...11Y}. As a consistency check, we also analyse the Subaru, R-band data presented in \citet[][]{2015ApJ...807L...2K} by applying our method to the galaxies in common to compare the recovered structural parameters (using our methodology) with those presented in \citet[][]{2016ApJS..225...11Y}.

\section{Spectral analysis}
\label{analysis}

We characterise the stellar kinematics of our targets by applying the penalized pixel fitting code \citep[{\tt pPXF};][]{2004PASP..116..138C, 2011MNRAS.413..813C} to the three spectra previously described (``W'', ``L'', and ``R''). The model stellar templates used in this and the subsequent steps are part of the new set of MILES\footnote{The models are publicly available at \url{http://miles.iac.es} and are based on the MILES empirical library \citep[][]{2006MNRAS.371..703S, 2011A&A...532A..95F}.} models (BASE models, following the MILES stars chemical pattern) with the BaSTI \citep[][]{2004ApJ...612..168P} isochrones \citep[][]{2015MNRAS.449.1177V, 2016MNRAS.463.3409V}. The spectral resolution of the data (FWHM $\sim$ 4.5 \AA) is translated into a minimum threshold below which we cannot reliably measure velocity dispersion values ($\sigma_{\rm min}$ $\sim$ 115 km/s). After convolving the models according to the instrumental dispersion, {\tt pPXF} preferred the lowest possible values for $\sigma$, suggesting that, taking into account the limited quality of the analysed spectra, the determination of the real stellar velocity dispersion of our UDG candidates is limited by our spectral resolution. Thus, the stellar velocity dispersion in all cases should be below $\sim$ 115 km/s.

\begin{figure*}
\centering 
\includegraphics[width = 0.93\textwidth]{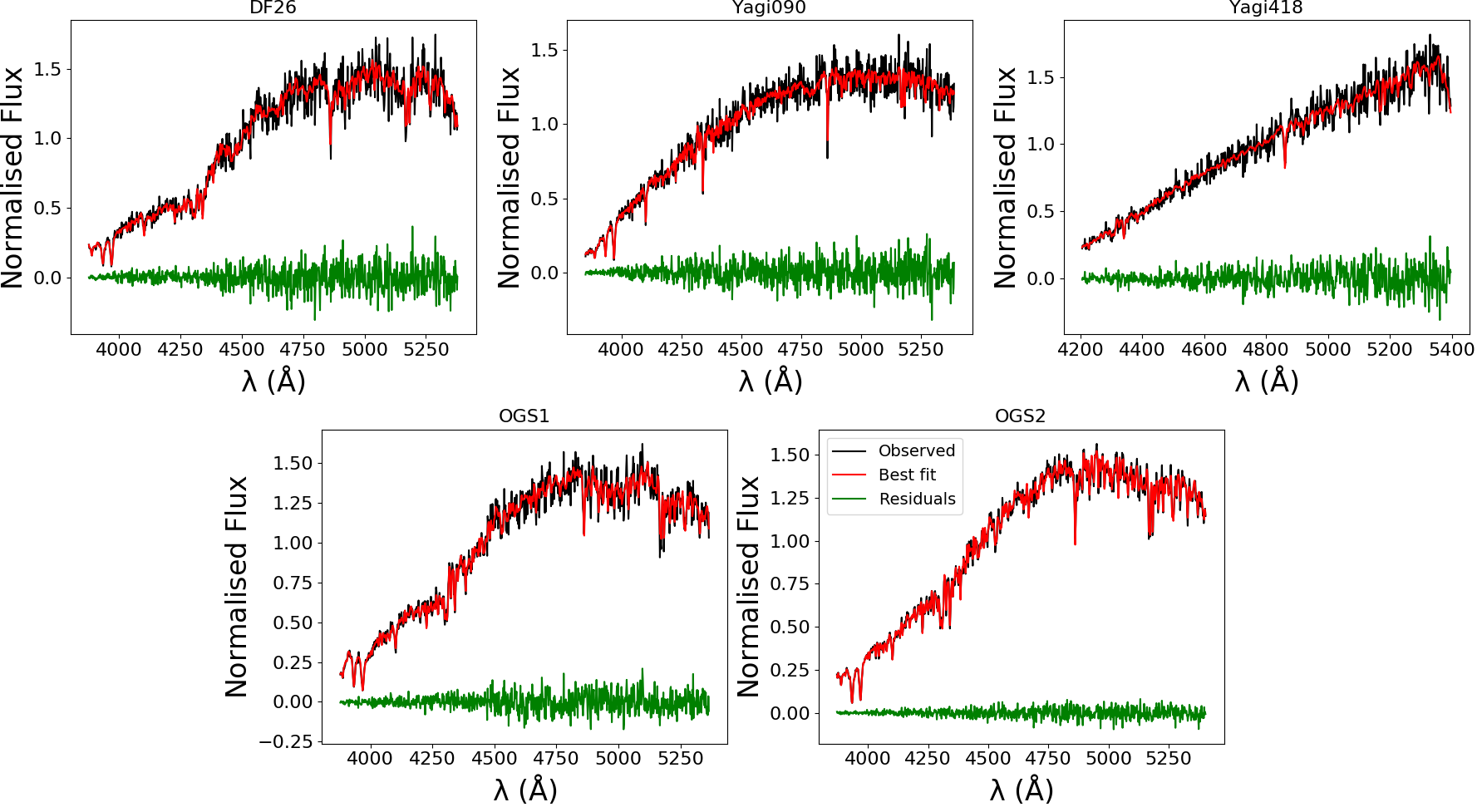} \\   
\caption{Normalised {\tt STECKMAP} spectral fits. All the spectra are at rest-frame. Observed spectra are represented by black lines while the best fits are in red. Residuals are shown in green.} 
\label{fits_plot} 
\end{figure*}

We study the stellar content of our UDG candidates using a methodology that combines different codes and that has been widely used and tested \citep[][]{2011MNRAS.415..709S, 2014MNRAS.437.1534S, 2015MNRAS.446.2837S, 2016MNRAS.456L..35R, 2017A&A...604A...4R, 2018Natur.553..307M}. In particular, this method has been proven unique at replicating a variety of SFHs from the, in principle, more reliable Colour-Magnitude Diagram fitting method applied to resolved stellar populations. \citet[][]{2015A&A...583A..60R} presented the first of such comparisons targeting a region within the Large Magellanic Cloud. This region has experienced star forming episodes since the formation of the system together with a continuous chemical enrichment that has been more pronounced in the last $\sim$ 3 Gyr, when the star formation rate has also increased. As a follow-up to this study, in \citet[][]{2018arXiv180504323R}, the authors extend this comparison to Leo~A, an extremely low-metallicity and young system characterised by an almost complete absence of stars older than 8 Gyr \citep[][]{2007ApJ...659L..17C}. The agreement between both approaches so far in these two systems further supports the results presented in this paper.

We apply this method only to the spectrum computed using all the brightest pixels (``W''). In short, after the stellar kinematics is determined and the best stellar model spectrum is found by {\tt pPXF}, we make use of {\tt GANDALF} \citep[Gas AND Absorption Line Fitting;][]{2006MNRAS.366.1151S, 2006MNRAS.369..529F} to check if our spectra are affected by any gaseous emission contribution in order to remove it. No gaseous emission was detected by {\tt GANDALF} in any of the 5 analysed galaxies. Finally, we use the STEllar Content and Kinematics via Maximum A Posteriori likelihood \citep[{\tt STECKMAP};][]{2006MNRAS.365...74O, 2006MNRAS.365...46O} code to properly characterised the stellar content shaping the observed spectra. Although {\tt STECKMAP} allows for a simultaneous stellar kinematics recovery, we decided to fix it to the {\tt pPXF} values and focus on the determination of the SFH \citep[minimising in this way the previously reported metallicity - velocity dispersion degeneracy, see][]{2011MNRAS.415..709S}.

Usually the results from {\tt STECKMAP} might slightly depend on the choice of input parameters, especially the so-called smoothing parameters, responsible for the smoothness of the final solution (regularization). To take that into account, the SFHs shown in this paper are the outcome of averaging the results of 12 different tests with different smoothing parameters (ranging from 0.001 to 1000) including errors (computed via 25 Monte Carlo simulations). In this way, the mass fraction at each age bin would be the median of all the solutions at that age bin and the errors would also be the median of all the errors. Similarly, the values given in Table~\ref{results_tab} (for the stellar age and metallicity) are the median of all the tests and the errors are computed as the percentile 5 for the lower limit and the percentile 95 for the upper one. We think that, given the subtle discrepancies among tests and the similar quality of the fits, a solution computed in this way reflects better the situation, not biasing the results towards any particular (somehow arbitrary) set of input parameters \citep[see][]{2015A&A...583A..60R, 2018arXiv180504323R}. Figure~\ref{fits_plot} shows the observed spectra (black) as well as the {\tt STECKMAP} fits (red) for the 5 UDG candidates analysed. Residuals are shown in green. The outcome of this analysis will be discussed in the next sections.

\section{UDG's stellar content}
\label{stars}

\begin{table*}
{
\scriptsize
\centering
\begin{tabular}{clcccccccccc}
\hline\hline
Galaxy         & Recession velocity &  z     &  R$_{\rm eff}$   & $\rm \langle log(Age_{LW}[yr]) \rangle$ & $\rm \langle [M/H]_{LW} \rangle$ & t$_{50}$ &  t$_{80}$ & t$_{90}$ & $M_{\star}$ & [Mg/Fe]  \\ 
               &  (km/s)  &          &   (kpc)    &   (dex)   &  (dex)                         & (Gyr)                            &  (Gyr)   & (Gyr)  & (M$_\odot$) & (dex)  \\ 
$(1)$          & $(2)$    & $(3)$   & $(4)$        &    $(5)$                & $(6)$                            &  $(7)$       &   $(8)$  & $(9)$ & $(10)$  & $(11)$  \\ \hline
 DF26          &  W: 6548.7 $\pm$ 27.1    &   0.02184  & 3.9 &  9.83$\substack{+0.08 \\ -0.09}$ (6.8 Gyr)   &  -0.78$\substack{+0.08 \\ -0.08}$    &  12.1  & 8.9 &  6.5   & 3.7$\substack{+0.2 \\ -0.2}\times$10$^{8}$ & $\sim$ 0.25 \\
               &  L: 6543.7 $\pm$ 24.4 (-5.0)   &                 &    &                   &                     &                  &    &   &   &   \\
               &  R: 6556.2 $\pm$ 29.2 (+7.5)   &                   &    &                   &                     &                  &    &   &  &   \\
 Yagi~090          &  W: 9420.1 $\pm$ 41.5    &   0.03142  & 2.0  &  9.76$\substack{+0.12 \\ -0.11}$ (5.8 Gyr)   &  --1.35$\substack{+0.05 \\ -0.04}$  &  9.4  & 7.1 &  5.8  & 1.0$\substack{+0.1 \\ -0.1}\times$10$^{8}$  &  $>$ 0.4 \\
               &  L: 9412.4 $\pm$ 45.1 (-7.7)   &            &     &                       &                     &                   &   &   &   &   \\
               &  R: 9438.7 $\pm$ 53.0 (+18.6)   &            &     &                     &                     &                   &   &    &   &   \\ 
 Yagi~418          &  W: 8190.6 $\pm$ 40.8    &   0.02732   &  1.6  &  9.91$\substack{+0.05 \\ -0.07}$ (8.1 Gyr)   &  --1.25$\substack{+0.05 \\ -0.04}$  & 9.3  &  7.4   &  6.7  & 8.5$\substack{+0.5 \\ -0.5}\times$10$^{7}$   &  $\sim$ 0.2 \\
               &  L: 8197.5 $\pm$ 52.5 (+6.9)   &              &         &                    &                     &                 &    &    &    &   \\
               &  R: 8187.4 $\pm$ 38.1 (-3.2)   &              &         &                    &                     &                 &    &    &    &    \\ 
 OGS1        &  W: 6367.3 $\pm$ 19.2    &    0.02124   & 1.5   &  9.93$\substack{+0.07 \\ -0.07}$ (8.5 Gyr)   &  --0.53$\substack{+0.05 \\ -0.06}$   &  13.8  &  12.11   &  10.6  & 3.1$\substack{+0.3 \\ -0.3}\times$10$^{8}$   &  $\sim$ 0.60 \\
               &  L: 6364.8 $\pm$ 24.2   (-2.5) &             &      &                     &                     &                   &   &  &    &   \\
               &  R: 6361.5 $\pm$ 23.1   (-5.8) &            &      &                    &                     &                     &  &   &   &    \\ 
 OGS2        &  W: 6980.0 $\pm$ 18.9    &   0.02328  &  1.3  &  9.85$\substack{+0.03 \\ -0.03}$ (7.1 Gyr)   &  -0.56$\substack{+0.03 \\ -0.02}$   &  10.2 &  6.6 & 4.1  &  3.8$\substack{+0.4 \\ -0.3}\times$10$^{8}$  & $\sim$ 0.30 \\ 
               &  L: 6981.9 $\pm$ 22.5  (+1.9)  &             &   &                         &                     &                  &    &    &    &   \\
               &  R: 6985.3 $\pm$ 20.0   (+5.3) &            &   &                       &                     &                &      &     &    &   \\  \hline
\end{tabular}
\caption{Galaxy properties. (1) Name given in this work to each of the analysed galaxies; (2) Velocity measurements from {\tt pPXF} (see Sect.~\ref{analysis}), for each galaxy we show three values: first row, the velocity from the spectrum ``W'' ($V_{W}$); second row, the velocity from spectrum ``L'' ($V_{L}$); and third row, the velocity from spectrum ``R'' ($V_{R}$); in parenthesis we added the velocity difference of the ``L'' and ``R'' spectra with respect to the ``W'' values; (3) redshift from spectrum ``W''; (4) effective radius (kpc); (5) median light-weighted (LW) logarithm of the age (in years) value from the {\tt STECKMAP} fits; (6) median light-weighted (LW) metallicity value from the {\tt STECKMAP} fits; (7) to (9) t$_{50}$ (t$_{80}$, t$_{90}$), defined as the look-back time at which half (80\%, 90\%) of the stellar mass of the system was formed; (10) stellar mass using the OGS observations and the M/L ratio predictions from the MILES models given the derived age and metallicities; (11) [Mg/Fe] values as derived from the analysis of line-strength indices. Note that columns (5) and (6) are light-weighted quantities, and thus, biased towards younger ages, whereas columns (7) to (9) are mass fraction, more representative of the old component, bulk of the mass of the systems.} 
\label{results_tab}
}
\end{table*}

Table~\ref{results_tab} summarises the main stellar kinematics and properties extracted from the analysis of the observed integrated spectra for the sample of galaxies. In this section we focus on the results from the analysis of spectrum ``W'' (whole galaxy).

Given the projected clustercentric distance of the analysed targets ($\le$ 0.5 Mpc), objects with {\it cz} values ranging between 4000 and 10000 km/s can be accepted as members of Coma considering the virial mass from \citet[][]{2007ApJ...671.1466K} \citep[see also][]{2018arXiv180109686A}. Thus, the stellar kinematics measurements confirm Coma cluster membership of the 5 galaxies under analysis in this work. Confirmation of a Coma cluster association allows us to place these objects on a physical scale, with sizes ranging from R$_{\rm eff} \sim 1.3 - 3.9$ kpc. The analysed sample comprises faint and extended galaxies spanning a range of sizes and surface brightness values.

\begin{figure*}
\centering 
\includegraphics[width = 0.75\textwidth]{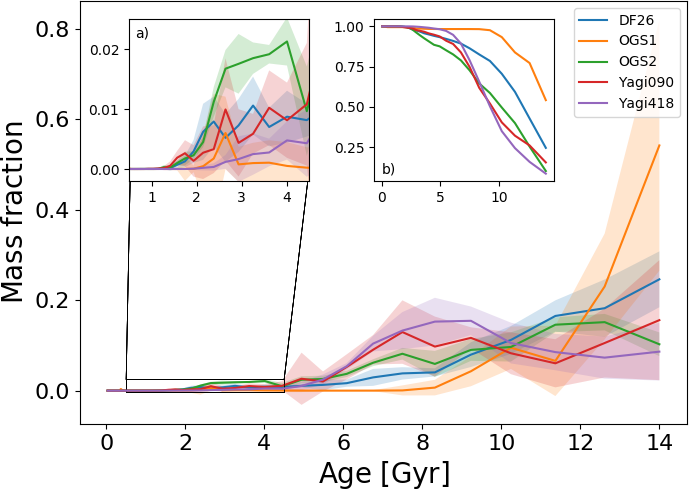} \\   
\caption{Star formation histories displayed by the sample of galaxies analysed. The derived star formation histories for all galaxies present a gradual decline until recently when a clear quenching of the star formation is observed. We show two different insets: a) Zoom at younger ages; b) cumulative mass fractions. The shaded areas represent the errors, computed taking into account all the tests (see text for details).} 
\label{sfh_plot} 
\end{figure*}

The SFHs for the five galaxies in Coma are shown in Fig.~\ref{sfh_plot}. All the galaxies seem to present similar overall SFH shapes, i.e. SFHs dominated by old stars, followed by a smooth decline with time of the star forming efficiency that end up with the absence of current star formation or stars younger than 1-2 Gyr. In Appendix~\ref{app_halfmass} we use mock spectra mimicking the emission of a stellar system with a given SFH using the MILES webtools\footnote{The MILES team provides users with a set of tools to fully exploit the models and to extract theoretical spectra with different SFH shapes \url{http://miles.iac.es/pages/webtools.php}.} to test to what extent we can constrain the real shape of the SFH given our age resolution and general uncertainties. To that aim we apply the same method as to the observed objects to the mock spectra after adding some noise to resemble the observed S/N. We conclude that we can rely on the recovered declining SFHs, however we cannot disentangle between a smooth or a bursty decline in the star formation. This uncertainty comes from the combination of the regularization used by {\tt STECKMAP} (preferring smooth solutions) and the struggling at discerning stellar properties at old ages (slow evolution, increase in the age-metallicity degeneracy, etc.). This implies that, at this stage, it is impossible to either confirm or invalidate the bursty star formation model proposed by \citet[][]{2017MNRAS.466L...1D}.

Despite this common behaviour, some subtleties are found among galaxies. For instance, while DF26, OGS1 and OGS2 best examplify the described behaviour, Yagi~418 and Yagi~090 display more extended SFHs as well as an apparent secondary peak in the mass fraction at ages $\sim$ 8-9 Gyr, probably a consequence of a delayed formation. In addition, the rate and time-scale at which the mass fraction decreases with time as well as the period of star forming inactivity slightly differ from galaxy to galaxy. 

The first SFH determination of UDGs were recently presented in \citet[][]{2018arXiv180109695F}. Although their cumulative mass fraction profiles (from {\tt STECKMAP}) resemble our derivations, the SFHs they show \citep[extracted using {\tt STARLIGHT},][]{2005MNRAS.358..363C} are drastically different from the ones shown in this work (even for the galaxies in common), being their SFHs bursty rather than extended and smoothly declining. However, we prefer our SFH determinations as: i) {\tt STARLIGHT} results rely in the quality of the (relative) flux-calibration of the studied spectra (always difficult in MOS data); ii) previous studies contrasting the {\tt STARLIGHT} and {\tt STECKMAP} performance as compared with the analysis of resolved stellar populations favour the usage of the latter \citep[][]{2015A&A...583A..60R}; and iii) despite the differences in the analysed spectra, the overall shapes of the cumulative mass fractions presented in both works agree qualitatively.

In order to further characterise the observed SFHs we compute the average stellar age and metallicity (see Table~\ref{results_tab}). In addition, other parameters have been measured to describe the shape of the observed SFH such as t$_{50}$, t$_{80}$, and t$_{90}$, being the look-back time (age of stars) at which 50\%, 80\%, and 90\% of the stellar mass of the galaxy was formed, respectively. All these parameters are in general agreement with previously derived values for UDGs \citep[][]{2017ApJ...838L..21K, 2017arXiv170907003G, 2018arXiv180109695F} as well as with a parallel analysis of the line strength indices.

To date, there are only 10 spectroscopic determinations of the stellar age and metallicity of UDGs, which makes the present work an important step forward in this field. \citet[][stacking the spectra of 4 UDGs in the Coma cluster]{2017ApJ...838L..21K} and \citet[][analysing separately spectroscopic data for 3 galaxies]{2017arXiv170907003G} find [Fe/H] values of $\le$ --1.5 and $\sim$ --1.1, respectively; as well as results that are consistent with old stars inhabiting the analysed systems. Special attention has to be paid to the recent results by \citet[][]{2018arXiv180109695F}, where the authors duplicate the number of spectroscopic UDG stellar age and metallicity determinations. Overall, they also find that UDGs are composed by intermediate-to-old, metal-poor, and slightly $\alpha$-enhanced stars. Two of their galaxies are included in the sample of galaxies analysed in this work (DF26 and Yagi~418). In these 2 cases, we obtain average age and metallicity values consistent with \citet[][]{2018arXiv180109695F}, and the wider wavelength range and the higher S/N of our OSIRIS data allow us to reduce the uncertainties on these measurements. In particular, we seem to find younger average ages (DF26: 6.8 $\pm$ 1.3 Gyr as compared to 7.88 $\pm$ 1.76 Gyr; Yagi~814: 8.1 $\pm$ 0.8 Gyr compared to 9.69 $\pm$ 2.02 Gyr), and slightly lower metallicities (DF26: --0.78 $\pm$ 0.08 as compared to --0.72 $\pm$ 0.18; Yagi~814: --1.25 $\pm$ 0.05 compared to --1.10 $\pm$ 0.95). 

\begin{figure}
\centering 
\includegraphics[width = 0.45\textwidth]{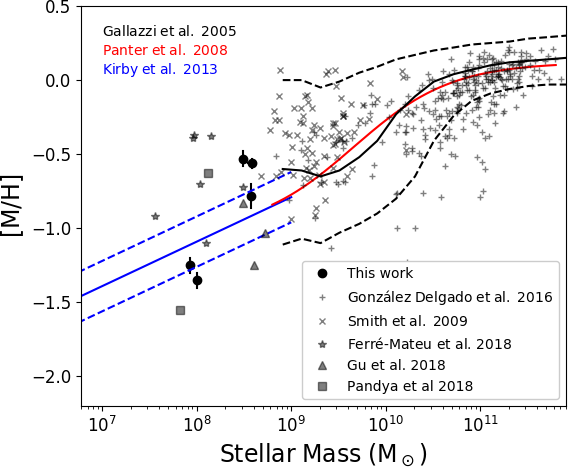} \\   
\caption{Stellar mass-metallicity relation for the galaxies under analysis (black circles). The different symbols  indicate the position on the mass-metallicity plane of different galaxies, from UDGs to massive galaxies \citep[][]{2015A&A...581A.103G}, including dwarf galaxies in Coma \citep[][]{2009MNRAS.392.1265S}. Overlaid are also the empirical relations for high-mass galaxies \citep[][]{2005MNRAS.362...41G, 2008MNRAS.391.1117P} and low-mass systems \citep[][]{2013ApJ...779..102K}.}
\label{MMR} 
\end{figure}

Despite the low number of UDGs to find a mass-metallicity relation for this kind of galaxies, \citet[][]{2018arXiv180109695F} found that UDGs seem to be more metal-rich than what expected for their masses \citep[][]{2013ApJ...779..102K}, apparently following the low-mass end of the relation for more massive systems \citep[][]{2005MNRAS.362...41G, 2008MNRAS.391.1117P}. Figure~\ref{MMR} places our galaxies in the stellar mass-metallicity plane along with all previous measurements for UDG galaxies as well as other galaxies \citep[][]{2009MNRAS.392.1265S, 2015A&A...581A.103G} and studied relations \citep[][]{2005MNRAS.362...41G, 2008MNRAS.391.1117P, 2013ApJ...779..102K}. In the case of the \citet[][]{2009MNRAS.392.1265S} dwarf systems we compute the stellar mass from their luminosities and considering M/L of single stellar populations of the measured ages and metallicities. We find that our systems encompass the relation derived for low-mass galaxies \citep[][]{2013ApJ...779..102K}, with the low metallicity ones being compatible with it. For the galaxies with the higher metallicities (DF26, OGS1 and OGS2), they seem to obey the \citet[][]{2005MNRAS.362...41G} relation or the Coma dwarf galaxies behaviour extrapolated to low stellar mass \citep[in agreement to what claimed in][]{2018arXiv180109695F}.

\begin{figure}
\centering 
\includegraphics[width = 0.45\textwidth]{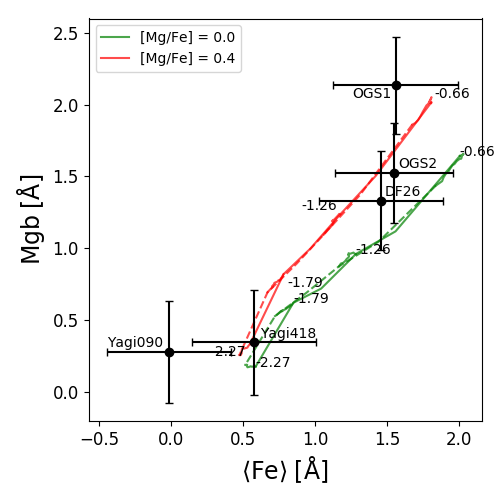} \\   
\caption{Mgb versus $\langle$Fe$\rangle$ diagram for the galaxies under analysis. The green and red overlaid grids correspond to the predictions from the scaled-solar and $\alpha$-enhanced MILES models \citep[][]{2015MNRAS.449.1177V}, respectively. The different ages and metallicities are chosen according to the values derived from {\tt STECKMAP}, i.e. 6 (dashed) and 14 (solid) Gyr and [M/H] = [--2.27, --1.79, --1.26, --0.66].} 
\label{indices} 
\end{figure}

Figure~\ref{indices} shows the Mgb versus $\langle$Fe$\rangle$ line-strength indices diagram \citep[LIS8.4,][]{2010MNRAS.404.1639V} as an attempt to constrain the [Mg/Fe] of our galaxies. Along with the observed line-strength measurements we overlaid the predictions from the MILES models \citep[][]{2015MNRAS.449.1177V} for the scaled-solar (green) and $\alpha$-enhanced (red) spectra ranging stellar ages and metallicities compatible to the derived values ({\tt STECKMAP}). We see that the metallicities are consistent with the {\tt STECKMAP}-derived ones and that the galaxies tend to be located towards the $\alpha$-enhanced part of the diagram. This apparent $\alpha$-enhancement is consistent with the fact that {\tt STECKMAP}, using the BASE MILES models, tends to reproduce weaker magnesium features than those observed in our targets, especially for OGS1, the UDG that seems to be more $\alpha$-enhanced (see Fig.~\ref{fits_plot}). Although our analysis finds (on average) slightly higher [Mg/Fe] values than those presented in \citet[][]{2018arXiv180109695F}, we found lower values for the galaxies in common (DF26: $\sim$ 0.25 as compared to 0.64 $\pm$ 0.25; Yagi~814: $\sim$ 0.2 compared to 0.27 $\pm$ 0.53). However, we should be cautious in our statements given the large error bars for the indices and the limited quality of UDG spectra. We also found a correlation between the stellar age of the systems and their [Mg/Fe] values in which old systems tend to display higher $\alpha$-enhancements. However, although this correlation is clear for DF26, OGS1 and OGS2 (the galaxies with the higher S/N spectra), we cannot draw any firm conclusions given the low number of galaxies under analysis.

\begin{figure*}
\centering 
\includegraphics[width = 0.95\textwidth]{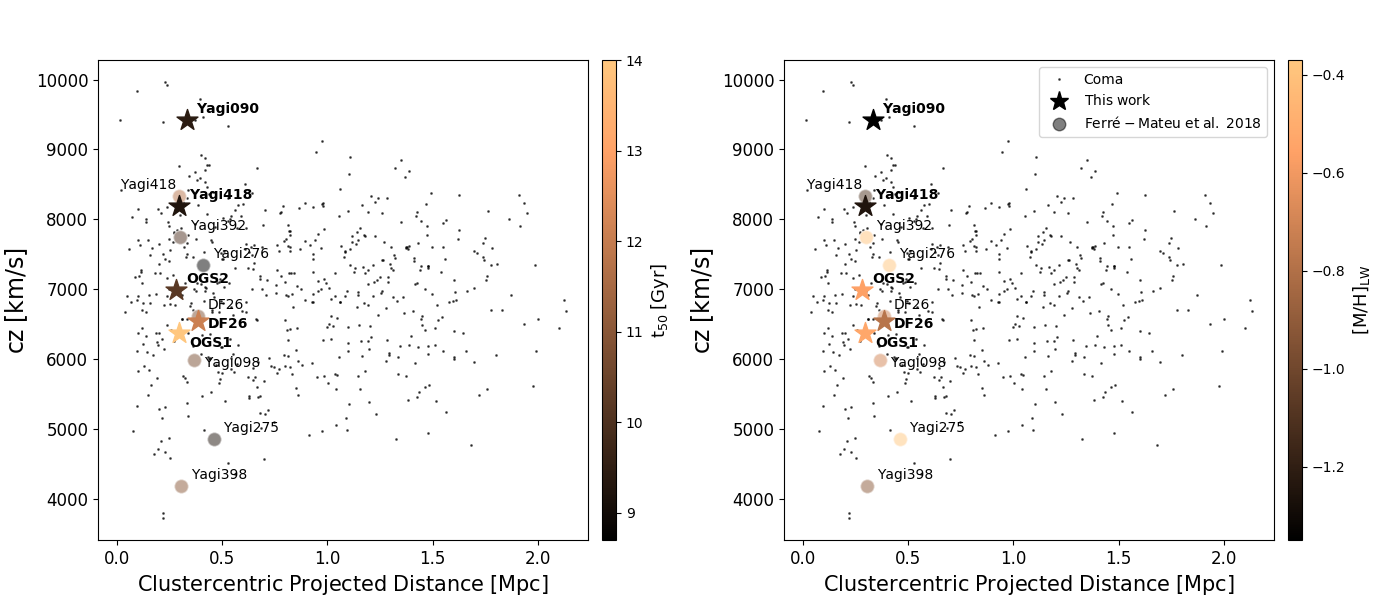} \\   
\caption{Phase-space diagram of the central region (2$^{\circ} \times$2$^{\circ}$) of the Coma cluster. Black dots represent galaxies around the cluster from SDSS DR12 \citep[][]{2015ApJS..219...12A}. The stars represent the galaxies analysed in this work and are colour-coded according to their t$_{50}$ (left panel) or their light-weighted metallicity ([M/H]$_{\rm LW}$, right panel). For completeness, we add the seven UDGs analysed in \citet[][]{2018arXiv180109695F} as circles with transparency. We assume a cosmological scale of 0.473 kpc/arcsec \citep[][]{2016ApJS..225...11Y}.} 
\label{caustic_plot} 
\end{figure*}

We also find that galaxies displaying recession velocities similar to the bulk of Coma \citep[cz = 7100 kms$^{-1}$,][]{2000ApJ...533..125K} tend to present larger values of t$_{50}$ than those with larger recession velocities (Yagi~418 and Yagi~090, see star symbols in Fig.~\ref{caustic_plot}). In addition, the former display higher stellar metallicities than the latter. This might be interpreted in terms of the in-falling history of galaxies towards the evolving Coma cluster, suggesting that galaxies located around the Coma redshift have been affected by the ever-changing Coma environment longer than those in the outskirts and that this has an impact on the evolution and properties of their stellar populations: inner galaxies formed quicker and earlier as well as underwent a quick chemical enrichment whereas outer-most galaxies have experienced a smoother evolution due to the lower degree of interactions with neighbour galaxies (see Sect.~\ref{discussion}). However, we should be cautious in these statements given the low number of galaxies under analysis and the fact that we are using projected clustercentric distances, ignoring the exact 3D location of our targets within Coma. We should highlight here the lack of other correlations with other observed properties such as t$_{80}$, t$_{90}$ or age. While a clear dependence of SFH with size (R$_{\rm eff}$) does not appear, it is intriguing to note that one of the smallest UDGs (OGS1) has almost no star formation after the first 6 Gyr. This is similar to what reported in \citet[][]{2017MNRAS.466L...1D} for the SFHs of the smallest galaxies in their sample (see their figure 4). In fact, the most compact systems are those displaying older ages.

\section[kin]{On the UDG's stellar rotation curve}
\label{kin}

One of the less-explored observables that could shed further light into the controversy regarding the nature of UDGs is their rotation curve. First \citet[][]{2017ApJ...836..191T}, and then \citet[][]{2017arXiv171006557S}, made use of single-dish, \hi~observations of a sample of blue UDGs to claim that they seem to be embedded in low-mass, high-spin haloes. The study of how UDGs rotate based on \hi~velocity widths provides an alternative approach to further delimit the possible dynamical mass of these systems \citep[see also][]{2015ApJ...798L..45V, 2016ApJ...830...23B, 2016ApJ...819L..20B, 2016ApJ...822L..31P}. The characterisation of the rotation curve shapes of these systems would provide a step forward. However, given the low surface brightness of UDGs, such measurements are extremely hard to obtain with current facilities. Nevertheless, even the most elementary approach can contribute to the understanding of these elusive systems. 

With the purpose of obtaining insight into the presence or absence of stellar rotation in these systems, we extracted two spectra at either side of the analysed galaxies avoiding their central regions (spectra ``L'' and ``R'', see Sect.~\ref{data}). The second column of Table~\ref{results_tab} shows the recovered velocities and uncertainties ({\tt pPXF}) for our sample of UDG candidates (V${_L}$~$\pm$~$\Delta V_L$ and V${_R}$~$\pm$~$\Delta V_R$). We define $\Delta V$ as $\rm |V_{L} - V_{R}|$. The comparison of this parameter with observed rotation curves will be used as a preliminary attempt to assess the possible rotation of these systems.

However, the uncertainties in the velocity determination of the 5 galaxies are not low enough as to properly obtain the value of $\Delta V$, finding a clear overlapping between both possible ranges of values (range$_L$ = [V${_L}$-$\Delta V_L$, V${_L}$+$\Delta V_L$] and range$_R$ = [V${_R}$-$\Delta V_R$, V${_R}$+$\Delta V_R$]). Nevertheless, and following a very conservative approach, we can safely establish an upper limit for such value in agreement with the errors, being $\Delta V_{\rm Max}$~=~Max([$|V{_L}+\Delta V_L - V{_R}-\Delta V_R$, $|V{_R}+\Delta V_R - V{_L}-\Delta V_L|$]). The values of $\Delta V_{\rm Max}$ that we recovered for DF26, Yagi~090, Yagi~418, OGS1, and OGS2 are 66.5, 124.4, 80.5, 50.3, and 45.9 km/s, respectively. With this approach we can be certain that the real $\Delta V$ value should be lower than $\Delta V_{\rm Max}$.

\begin{figure*}
\centering 
\includegraphics[width = 0.75\textwidth]{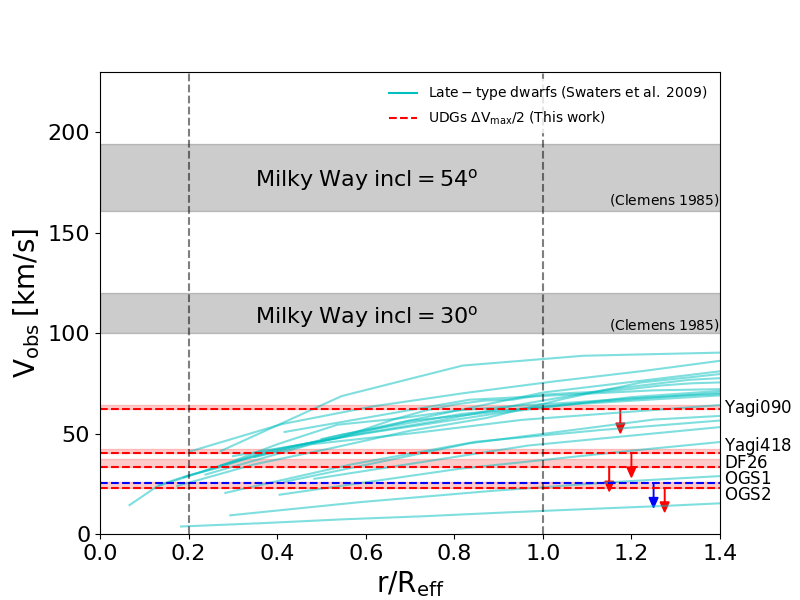} \\   
\caption{Analysis of the stellar rotation curves compatible with our velocity determinations. We compare with the Milky Way rotation curve from \citet[][]{1985ApJ...295..422C} mimicking the inclinations of our UDGs (30$^{\rm o}$ to 54$^{\rm o}$, grey areas) and with late-dwarf observed rotation curves \citep[][cyan lines]{2009A&A...493..871S}. The upper limit for the rotation curves ($\Delta V_{\rm Max}$/2) is shown by the red horizontal dashed lines. The vertical dashed lines indicate the mean position from where the spectra where extracted. Possible uncertainties induced by the inclination errors are given by the red shaded areas. OGS1 is the only example for which the slit was clearly not aligned with the major axis of the galaxy, hampering a proper $\Delta V_{\rm Max}$ computation (blue dashed line). Arrows are added to emphasize the fact that these values are upper limits.} 
\label{fig:rotation_tests} 
\end{figure*}

Figure~\ref{fig:rotation_tests} compares the $\Delta V_{\rm Max}$/2 values for the UDG candidates (red dashed lines) to the values that would be observed for the Milky Way \citep[][]{1985ApJ...295..422C} if observed with the inclination of our targets (30$^{\rm o}$ to 54$^{\rm o}$, gray areas). On the other hand, it also compares with one of the most complete samples of \hi~rotation curves of dwarf galaxies \citep[][cyan dashed lines]{2009A&A...493..871S}. In particular, we restrict our comparison to the galaxies from their high-quality subsample, i.e. those whose rotation curves are flagged as reliable \citep[see also][]{2014MNRAS.439..284R}. In addition, we make use of the rotation curve parametrizations provided in \citet[][from H$\alpha$ velocity maps]{2006ApJ...640..751C} to assess, via 100 Monte Carlo simulations, our errors in the $\Delta V_{\rm Max}$ due to uncertainties in the slit position angle (generally aligned with the major axis of the galaxies except for OGS1, blue dashed line) and the galaxy inclination (see red-shaded areas). 

This analysis can be considered as the first attempt at characterising the stellar rotation curves of UDGs. We can clearly rule out a Milky Way rotation and we show the consistency of all our measurements with the typical rotation curves observed in dwarf systems. However, we are only establishing upper limits for $\Delta V_{\rm Max}$, and thus, an absence of rotation in these systems cannot be discarded. Also, note that we are not able to determine the velocity dispersion of these systems due to the limited spectral resolution of OSIRIS. However, we can establish an upper limit. UDGs display velocity dispersion below $\sim$ 115 km/s in all cases\footnote{Following a similar approach to that presented in \citet[][]{2016ApJ...819L..20B} to estimate halo masses from velocity dispersions, our spectral resolution translates into an upper limit to the halo masses of $\sim$~10$^{12}$--10$^{13}$ M$_{\odot}$, which would be typical of Milky Way and more massive systems. Direct measurements of the stellar velocity dispersion at higher spectral resolution would be very useful to better constrain this quantity.}. Deep analysis with larger spectral and spatial resolution are needed to properly describe the kinematics of UDGs.

\section{Discussion}
\label{discussion}

In the last few years, since the renewed interest on these systems by \citet[][]{2015ApJ...798L..45V}, the task of disentangling the origin of Ultra Diffuse Galaxies has received considerable attention. While simulations are proposing a number of evolving pathways as how these galaxies acquired the observed characteristics, observations struggle to gather the high-quality data needed to study their dynamical and stellar population properties. This paper is another step in the attempt of discerning between the two main proposed formation scenarios of UDGs, namely that UDGs are 'failed' giant galaxies \citep[][]{2015ApJ...798L..45V, 2016ApJ...822L..31P, 2017ApJ...844L..11V} or peculiar, extended dwarf systems \citep[][]{2016MNRAS.459L..51A, 2017MNRAS.466L...1D}. In this paper, we thoroughly analyse OSIRIS@GTC spectroscopic data for a sample of 5 UDG candidates. We confirm Coma membership for all of them and find that 4 (DF26, Yagi~090, Yagi~418 and OGS1) are clear examples of UDGs (using the \citealt[][]{2015ApJ...798L..45V} size criterion), whereas OGS2 might be better considered an extended, low-surface brightness dwarf galaxy. All these galaxies are characterised by  old (light-weighted age of $\sim$ 7 Gyr), metal-poor (given their stellar masses, [M/H] $\sim$ -1.0), $\alpha$-enhanced ([Mg/Fe] $\sim$ 0.4) stellar populations in agreement with several previous studies of UDGs using photometric and spectroscopic data (see Table~\ref{results_tab}).

One of the most complete characterisations of the mass assembly time scales of nearby galaxies was recently published by \citet[][]{2017A&A...608A..27G} using CALIFA data \citep[][]{2012A&A...538A...8S}. They analyse the stellar content of 661 galaxies with stellar masses ranging from 10$^{8.4}$ to 10$^{12}$~M$_\odot$ including ellipticals to Sd galaxies. The cumulative mass fraction profiles and t$_{80}$ values displayed by our sample of UDGs are more compatible with those galaxies within their lowest stellar mass bins (10$^{8.4}$ to 10$^{9.9}$ M$_\odot$) and lowest stellar mass surface density (10$^{1.2}$ to 10$^{2.0}$ M$_\odot$/pc$^2$). The main differences are found at the youngest ages, where we find clear signs of a quenching in the star formation that they do not detect. However, we should bear in mind that less massive galaxies (as our sample) are more susceptible to external influences than more massive systems \citep[][]{2017MNRAS.470..815S}. Since the CALIFA galaxies are mainly located in low density environments \citep[][]{2016A&A...594A..36S} and the UDGs analysed in this work lie in the Coma cluster, we may relate the existence of this quenching to either environmental effects, or the internal properties of the UDGs under analysis. Based on this comparison with the SFHs of CALIFA galaxies we find that the recovered SFHs seem to be more consistent with those of the lowest mass systems. Unfortunately, the \citet[][]{2017A&A...608A..27G} sample lacks of galaxies in the precise stellar mass range covered in this work. A direct comparison with the SFHs of dwarf galaxies is needed.

\citet[][]{2009MNRAS.396.2133K} showed the SFHs for a sample of 16 dwarf ellipticals in the Fornax cluster and other groups. They found that their galaxies are mainly dominated by an old component (coeval with that of massive ellipticals) and an extended component towards younger ages (in some cases even till recent times) with the presence also of some intermediate-age population. In a more recent work, \citet[][]{2015MNRAS.452.1888R} reached similar conclusions. Their 12 dwarf ellipticals (10 in the Virgo cluster and 2 in the field) display star formation extending until a few Gyr ago while dominated (especially in mass) by an old component (generally older than 10 Gyr). In addition, \citet[][]{2009MNRAS.392.1265S} present a thorough characterisation of the stellar content in 89 dwarf galaxies in the Coma cluster. Their recovered average ages and metallicities are in the range of our values, although they also find some dwarf systems with younger ages especially located in the periphery of the cluster, where the conditions are significantly different than those in the region analysed in this work. Apart from that, they also show cumulative mass fractions displaying a clear absence of young stars (2-3 Gyr old) although in their case the decline of the mass fraction towards younger stars seem to be smoother as the one recovered here.

However, not all dwarf galaxies display such similar SFHs. \citet[][]{2015ApJ...811L..18G}, analysing resolved stellar populations of a sample of Local Group dwarf galaxies, distinguished between {\it slow} and {\it fast} dwarfs depending on the shape of their SFHs. Those classified as {\it slow} display extended SFHs until the present whereas {\it fast} dwarfs are mainly dominated by an initial and short formation event with no star formation during the last $\sim$ 8-9 Gyr. The authors claim that this dichotomy should be imprinted in the early stages of formation of the systems (initial environmental conditions) rather than driven by more recent environmental effects. Nevertheless, the later can also affect the final shape of the observed SFH, especially for the {\it slow} dwarfs. In the line of their reasoning, the SFHs here-derived have characteristics in common with both, {\it slow} and {\it fast} dwarfs, with our systems being more dominated by older populations than {\it slow} dwarfs and more extended than {\it fast} dwarfs. In particular, we should highlight the fact that the galaxies with more extended SFHs and important contribution of intermediate populations are Yagi~418 and Yagi~090, those located in the periphery of the coma cluster (see Fig.~\ref{caustic_plot}), in agreement with the scenario proposed in \citet[][]{2015ApJ...811L..18G}.

In this sense, it is worth noting the case of DDO~44, a local UDG analogue. DDO~44 is a low surface brightness ($\rm \langle \mu_{B} \rangle \sim $ 24.1 mag/arcsec$^2$), metal-poor ([Fe/H] = --1.7) dwarf spheroidal with an effective radius of $\sim$ 1 kpc, located at 3 Mpc and associated with NGC~2403 \citep[group environment, see][]{1999A&A...352..399K}. The proximity of this system allows for careful analyses of individual stars as well as observations of stellar colour-magnitude diagrams. Such analyses reveal that DDO~44 displays an extended SFH, being dominated by an old stellar population with the presence of some intermediate component \citep[][]{2006PASP..118..580A, 2010ApJ...724.1030G}, a SFH that is very similar to those derived in this work.

Furthermore, the [Mg/Fe] ratios of the UDGs provide an important constraint (although note the significant uncertainties of these derivations presented in Sect.~\ref{stars}). The UDGs analysed in this work seem to display large values of [Mg/Fe] ($\sim$ 0.4 on average). If we compare them with typical [Mg/Fe] values for dwarf systems we see that they are more similar to the values recovered in the outskirts of dwarfs \citep[][]{2018MNRAS.tmp..511S} or those of quenched dwarfs \citep[][]{2017MNRAS.470..815S}. In addition, also the mass-metallicity relation (see Fig.~\ref{MMR}) displayed by our UDGs seems to be more in agreement with Coma dwarf galaxies \citep[][]{2009MNRAS.392.1265S} or dwarf galaxies in general \citep[][]{2013ApJ...779..102K}.

Therefore, all the evidences seem to suggest that the stellar populations of the analysed galaxies, irrespective of their physical size, are very similar and are in agreement with those found in (quenched) dwarf systems. We note, of course, that in the above discussion one should bear in mind our small sample size and the fact that the detailed SFHs  will depend on each galaxy's own particular evolutionary history.

Apart from this characterisation of the stellar content, in Sect.~\ref{kin} we show a first tentative attempt to describe the stellar rotation curve of UDGs and compare it with the Milky Way rotation curve and with typical rotation curves of dwarf galaxies. Despite the amount of simplifications (not considering asymmetric drift, having just two points, etc.) and the extremely conservative approach followed, the results presented there tempt us to claim that, if UDGs rotate (option that cannot be discarded with this analysis), they should display a dwarf-like rotation rather than a massive system rotation \citep[see also][]{2017ApJ...836..191T, 2017arXiv171006557S}. Another similarity with systems embedded in dwarf-like haloes.

Thus, the spectroscopic analysis presented in this work could be in agreement with an evolution pathway in which the analysed UDGs (within the Coma cluster) might be dwarf galaxies that formed in isolation or at least in a less-dense environment and progressively fall into the cluster potential causing a gradual decrease in the star forming efficiency until it is halted \citep[e.g.][]{1986ApJ...303...39D, 1999ApJ...513..142M, 2010MNRAS.402.1599S, 2010AdAst2010E..25M}. However, we cannot rule out the effect of other (internal or external processes) shaping the final properties of UDGs. In addition, we do not find clear differences in the dynamical or stellar populations properties between confirmed UDGs and other dwarf galaxies. But, how do our findings link with proposed scenarios on the origin of UDGs? 

On the one hand, \citet[][]{2015ApJ...798L..45V} first suggested that UDGs might be ``failed'' Milky Way-like galaxies that could not fully develop its massive stellar disc. In that line, \citet[][]{2015MNRAS.452..937Y} formulate a model in which late-type spiral galaxies embedded in $\sim$ 10$^{11}$ M$_{\odot}$ dark matter haloes could experience an early and rapid quenching (z $\sim$ 2) as they are accreted towards over-dense environments. However, we do not find any hints for such early quenching on the star formation of the analysed galaxies, but rather a smooth decline in the star forming efficiency that is not fully quenched until the last $\sim$ 2 Gyr. This, added to the rest of evidences presented in this work favouring a dwarf-like nature of UDGs indicates that another explanation would be preferred.

The other main scenario to explain the observed properties of UDGs suggests that they are genuine dwarf galaxies either embedded in high angular momentum dwarf haloes \citep[][]{2016MNRAS.459L..51A, 2017MNRAS.470.4231R} or that had experienced feedback-driven gas outflows due to a bursty star formation history \citep[][]{2017MNRAS.466L...1D, 2017arXiv171104788C}. This work has to be added to the myriad of observational evidences favouring an internal processes scenario \citep[][]{2016ApJ...830...23B, 2016A&A...590A..20V, 2017MNRAS.468..703R, 2017MNRAS.468.4039R, 2017A&A...601L..10P, 2017ApJ...836..191T, 2018arXiv180109695F}. However, we cannot disentangle between the two possible alternatives. High spinning dark matter haloes do not necessarily imply highly rotating stellar systems (which we demonstrate UDGs are not), so this analysis cannot discard or favour the first alternative. In the same sense, the recovered SFHs are compatible with a series of bursty star forming episodes declining with time (favouring the stellar feedback origin), yet, it is also compatible with just a gradual decrease in the star formation efficiency.

Finally, we also need to take into account the trends displayed by t$_{50}$ and the stellar metallicity according to the position of the galaxies within the Coma cluster (see Fig.~\ref{caustic_plot}). According to our findings, galaxies that are currently located in the outskirts of Coma have lower stellar metallicities and their SFH are more extended. \citet[][]{2017MNRAS.468.4039R} already suggested a possible scenario in which blue low surface brightness galaxies born in the field can acquire (red) UDG characteristics after being processed in a higher density environment (quenching and subsequent passive evolution). Similar results are obtained by \citet[][]{2017arXiv170904474G} where the authors compare the characteristics of their blue and red low surface brightness galaxies \citep[using the Hyper Suprime-Cam Subaru Strategic Program,][]{2018PASJ...70S...4A}. They find that their "red" sample displays redder colours, fainter magnitudes, more regular shapes and are more centrally concentrated around clusters than their "blue" counterpart. The results presented in this paper can be interpreted in a similar fashion. The higher metallicities and less extended SFHs of galaxies at the redshift of the bulk of Coma can be consequence of an earlier accretion (or even formation in the cluster) to the Coma cluster inducing star formation at older ages and a quicker chemical enrichment. On the other hand, those galaxies nowadays located at higher redshift (periphery) have experienced a slightly smoother process (due to a formation in a slightly less dense environment) resulting in lower metallicities and more extended SFHs \citep[again consistent with the scenario proposed in][]{2015ApJ...811L..18G}. The inclusion of the galaxies from \citet[][]{2018arXiv180109695F} further support this claim. Also their galaxies located towards the closer end of the range of redshifts tend to display lower values of the stellar metallicities and t$_{50}$.

One of the remaining details is the lack of a clear correlation between t$_{90}$ (proxy for the quenching time) and any other properties. However, we must bear in mind the limited number of objects under analysis and the peculiarities that might arise for individual systems (initial and environmental conditions, gas reservoir, etc.). In that sense, \citet[][]{2017arXiv171104788C} predict that different levels of feedback in galaxies embedded in normal spinning dwarf haloes might result in a wide range of quenching times. The authors show also some correlations between this quenching time and other properties such as stellar mass, age or metallicity. Although we do not find such correlations, we cannot discard the existence of such relations due to the low number of galaxies in this work.

\section{Conclusions}
\label{conclusions}

In this paper we present one of the most complete and thorough characterisations of the stellar content from spectroscopic data in UDGs up to date. We confirm that the 5 analysed UDGs lie in the Coma cluster.
These UDGs, irrespective of their physical sizes, display similar SFHs characterised by old ($\sim$ 7 Gyr), metal-poor ([M/H] $\sim$ --1.0) and $\alpha$-enhanced ([Mg/Fe] $\sim$ 0.4) stars. In addition, the recovered SFHs as well as the tentative stellar rotation curve measurements presented in this work quantitatively agree with those found in dwarf galaxies (although we cannot discard an absence of rotation in them), favoring  a formation scenario based on internal processes. Thus, we suggest that the analysed UDGs in Coma are dwarf galaxies formed in low-density environments whose specific properties are shaped by the combined effect of a slow process of immersion in the Coma cluster and internal processes (high spin halo or bursty SFH). The possibility of finding field UDGs remains open, acquiring in this case the observed properties mainly via internal processes. We provide here new and clear evidence favouring the dwarf-like galaxy scenario. The extension of this kind of analysis to larger samples of galaxies is of critical importance to fully understand the formation and evolution of these systems. 

\section*{Acknowledgements}

The authors thank the anonymous reviewer for a careful reading of the manuscript and useful comments that have helped to improve previous versions of this paper. We thank Jin Koda for kindly sharing their Subaru R-band data of the Coma cluster presented in \citet[][]{2015ApJ...807L...2K} with us. This research has been mainly supported by the Spanish Ministry of Economy and Competitiveness (MINECO) under grant AYA2016-77237-C3-1-P. TRL also acknowledges support from grant AYA2014-56795-P from MINECO. M.A.B. acknowledges financial support from the Svero Ochoa Excellence programme (SEV-2015--0548). A.D.C. acknowledges financial support from a Marie Sk\l{}odowska-Curie Individual Fellowship grant, H2020-MSCA-IF-2016, Grant agreement 748213 DIGESTIVO. I.T.C. acknowledges financial support from the European Union’s Horizon 2020 research and innovation programme under Marie Sk\l{}odowska-Curie grant agreement No 721463 to the SUNDIAL ITN network.

Funding for Sloan Digital Sky Survey-III has been provided by the Alfred P. Sloan Foundation, the Participating Institutions, the National Science Foundation, and the U.S. Department of Energy Office of Science. The SDSS-III web site is http://www.sdss3.org/.

This research makes use of python (\url{http://www.python.org}); Matplotlib \citep[][]{hunter2007}, a suite of open-source python modules that provide a framework for creating scientific plots; and Astropy, a community-developed core Python package for Astronomy \citep[][]{astropy2013}.

\bibliography{bibliography}

\bsp

\appendix

\section{SFH recovery tests: On the reliability of the {\tt STECKMAP} results}
\label{app_halfmass}

The credibility of the {\tt STECKMAP} results in combination with the complete methodology used in this work has been extensively tested in the literature \citep[][]{2008MNRAS.385.1998K, 2011MNRAS.415..709S}. In particular, in \citet[][]{2015A&A...583A..60R} we prove the consistency between the {\tt STECKMAP} results and the analysis of deep colour-magnitude diagrams of resolved stars regarding the general shape of the time variation of the star formation rate and the chemical enrichment. However, the low surface brightness of the objects analysed in this work hampers the gathering of spectroscopic information of similar quality as that of the spectra analysed in those works. As a consequence we have used the MILES webtools to generate 42 mock spectra from parametric star formation histories. To mimic the quality of the analysed spectra we add some noise to these mock spectra and run {\tt STECKMAP} to assess the reliability of the SFHs shown in Fig.~\ref{sfh_plot}.

\begin{figure}
\centering 
\includegraphics[width = 0.4\textwidth]{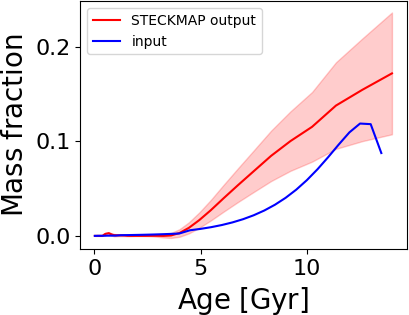} \\
\includegraphics[width = 0.4\textwidth]{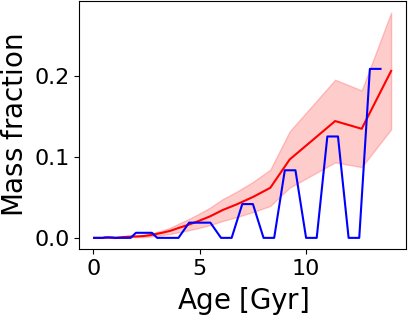} 
\caption{Star formation history recovery tests. The computed star formation histories are compatible with a continuously (top) or a bursty (bottom) declining star formation rate. In blue the input SFH from the MILES webtools, in red the recovery from STECKMAP (see text for details).} 
\label{app:SFH_tests} 
\end{figure}

The comparison between the shapes of the input SFHs and those recovered by {\tt STECKMAP} confirm the reliability of our measurements. {\tt STECKMAP} easily recovers the input SFH, especially in those cases of SFHs declining with time. Figure~\ref{app:SFH_tests} shows examples of the input SFH shapes from the MILES webtools (blue) that are compatible with the SFHs we recover for the analysed sample of UDG candidates. The {\tt STECKMAP} output (recovered SFH) is shown in red. As the figure suggests, {\tt STECKMAP} successfully reproduced the general declining trend, however it cannot disentangle between a smooth decline (top) or a bursty, episodic one (bottom); so, we must be aware of these limitations when drawing conclusions regarding the true nature of UDGs.

\begin{figure}
\centering 
\includegraphics[width = 0.48\textwidth]{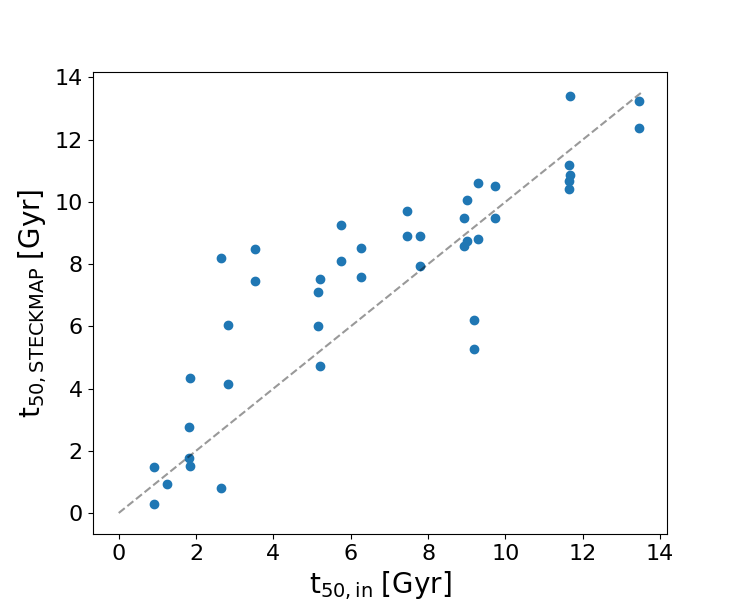} \\   
\caption{Recovery of the t$_{50}$ parameter. Blue points represent each of the 42 tests. The gray dashed line represents the one-to-one relation.} 
\label{app:HMT} 
\end{figure}

Apart from the general shape of the SFH, in the main body of the paper we make use of other parameters such as the t$_{50}$ to demonstrate that, not only the general shape is recovered but also the rate at which the stellar mass fraction decreases with time. Figure~\ref{app:HMT} compares the recovered (y-axis) and the input (x-axis) t$_{50}$ values from this comparison. These tests allow us to confirm the good performance of our methodology at disentangling SFHs from spectroscopic data, even in the low S/N regime.

\end{document}